 \documentclass[final,3p,times,twocolumn]{elsarticle}

\usepackage{amssymb}

\usepackage{dcolumn}                             
\usepackage{multirow}                   
\usepackage{rotating}
\usepackage{amsmath, amsthm, amsfonts}
\usepackage{fullpage}
\usepackage{times}
\usepackage{xcolor}
\usepackage{shadethm}
\usepackage{feynmp}
\usepackage{array}
\usepackage{booktabs}
\usepackage{float}

\journal{Physics Letters B}
\begin{document}

\begin{frontmatter}


\title{Quark-Meson Coupling model based upon the Nambu-Jona Lasinio model }


\author{D.~L.~Whittenbury}
\author{M.~E.~Carrillo-Serrano}
\author{A.~W.~Thomas}

\address{CSSM and ARC Centre of Excellence for Particle Physics at the 
Terascale,\\ School of Physical Sciences,
University of Adelaide,
  Adelaide SA 5005, Australia}

\begin{abstract}
The NJL model for the octet baryons, using proper time regularization to simulate 
some of the features of confinement, is solved self-consistently in nuclear 
matter. This provides an alternative framework to the MIT bag model which  
has been used in the quark-meson coupling model. After fitting the parameters 
of the model to the saturation properties of symmetric nuclear matter the 
model is used to explore the equation of state of pure neutron matter as well as 
nuclear matter at densities relevant to heavy ion collisions. With a view 
to future studies of high mass neutron stars, the binding of hyperons is also 
explored.
\end{abstract}


\end{frontmatter}


\section{Introduction}
\label{sec:intro}
There is little doubt that Quantum Chromodynamics (QCD) is the 
correct theory of the strong interaction. 
However, the issue of connecting this more fundamental theory to 
traditional nuclear physics is extremely challenging. 
Of course, we have hints of what might be involved 
through phenomena such as the 
EMC effect~\cite{Aubert:1983rq,Geesaman:1995yd} but we are far from a full 
understanding of the influence of quark degrees of freedom and 
their implications for the complex phenomena emerging from QCD.

The Quark-Meson Coupling (QMC) model goes beyond 
the majority of nuclear models by explicitly 
treating baryons as extended objects.
It is a relativistic quark level model which has been extensively used 
to study nuclear matter~\cite{Guichon:1987jp}, finite 
nuclei~\cite{Guichon:1995ue}
and neutron stars~\cite{RikovskaStone:2006ta}. 
The model has recently been shown 
to provide a remarkably accurate description of the ground-state 
properties of 
atomic nuclei across the periodic table, 
in terms of a derived, density-dependent 
effective NN potential~\cite{Stone:2016qmi}. Within QMC the MIT bag model is 
used as the model of hadron structure, although one need not restrict oneself 
to this. Indeed, it is clearly of interest to extend the approach to other 
models of hadron structure. For example, Bentz and Thomas ~\cite{Bentz:2001vc} 
were the first to develop such a theory by hadronising the NJL model, 
which embodies different aspects of QCD, 
notably spontaneous chiral symmetry breaking.
The aim of this letter is to consider the effect of hadron structure on 
nuclear matter properties within this complementary model.

Within the QMC model the in-medium changes of the baryon properties, 
such as masses, scalar couplings and so on, are calculated by self-consistently 
solving the bag equations, including the effect of the mean fields generated by 
other nucleons. The masses are then parametrised as functions of the mean 
scalar field as
\begin{equation}
M^*_B = M_B - w_{B}g_{\sigma N}\bar{\sigma}
+ \frac{d}{2}\tilde{w}_{B}\left(g_{\sigma N}\bar{\sigma}\right)^2 \ ,
\label{eq:massparam}
\end{equation}
(where the weightings $w_{\sigma B}$ and $\tilde{w}_{\sigma B}$ simply
allow us to express the density dependent couplings of the mean scalar field 
to each hadron in terms of the unique coupling to the nucleon in free 
space, $g_{\sigma N}$. Using this parametrisation and the  
corresponding density dependent coupling,  
we can solve for the equation of state in a manner analogous to the Walecka 
model~\cite{Chin:1977iz,Serot:1979cc,Serot:1984ey,Serot:1997xg}, that is at the hadronic
level. In this way the sub-structure of the baryons is entirely contained in the 
mass parametrisation. 
In Ref.~\cite{MyThesis,PhysRevC.89.065801,Whittenbury:2015ziz},  we used 
the bag model parametrisation given in Ref.~\cite{Guichon:2008zz}, 
which includes the effects of one gluon exchange. 
Here we present a new variation of the QMC model with a 
mass parametrisation obtained by solving the Faddeev equation
derived from the proper time regularised NJL model. 
Then, using this new mass parametrisation, we calculate the equations of state 
of Symmetric Nuclear Matter (SNM) and Pure Neutron Matter (PNM) 
in a Hartree-Fock approximation. 

In Ref.~\cite{PhysRevC.89.065801} we extended the QMC model by performing a 
Hartree\textendash Fock calculation including the full vertex structure for 
the vector mesons. This extension only alters the exchange contribution,
including not only the Dirac vector term, as was done in
Ref.~\cite{RikovskaStone:2006ta}, but also the Pauli tensor term. These
terms were already included within the QMC model by Krein~{\it et
  al.}~\cite{Krein:1998vc} for symmetric nuclear matter and more
recently by Miyatsu {\it et al.}~\cite{Miyatsu:2011bc}. 
We generalised the work of Krein~{\it et
  al.}~\cite{Krein:1998vc} by evaluating the full exchange 
terms for all octet baryons and
adding them, as additional contributions, to the energy density. A consequence
of this increased level of sophistication is that, if we insist on using the 
hyperon couplings predicted in the simple QMC model, with no meson coupling to
the strange quarks, the $\Lambda$ hyperon is no longer bound. 
Addressing the under-binding of the $\Lambda$ hyperons in nuclear matter 
and accounting for the known existence of $\Lambda$-hypernuclei 
without the need to phenomenologically rescale couplings is a pressing 
issue. As the scalar couplings are dependent on the model of hadron structure 
through Eq.~(\ref{eq:massparam}), it is interesting to consider an 
alternative to the conventionally used bag models.
%
 \begin{table*}[t] 
 \centering
 \resizebox{\textwidth}{!}{
 \begin{tabular}{ccccccccccccc}
  \toprule[1.5pt]
 & $m_{\ell}$ & $m_{s}$  & $M_{\ell}$ & $M_{s}$ & $\Lambda_{\rm IR}$& $\Lambda_{\rm UV}$ & $G_{\pi}$ & $Z_\pi$ & $Z_\pi(0)$ & \\
 & [MeV] & [MeV] & [MeV]& [MeV] & [MeV] & [MeV] & [GeV$^{-2}$] & & & \\
\midrule
 &  16.43   & 324.32   &  400 & 592.17 & 240 & 644.87  & 19.044 & 17.853 & 18.500 & \\
   \bottomrule[1.5pt]
 \end{tabular}
 }
 \caption{
Values of  the proper time regularised NJL model parameters. Tabulated are the current and constituent quark masses, infra-red and ultra-violet cut-offs, and  scalar and pion effective couplings (dimensionless) evaluated at $q^2 = m_\pi^2$ and $q^2 = 0$.  
 The parameter set used $M_{\ell}$, $\Lambda_{\rm IR}$, $f_{\pi}=93$~MeV and  $m_{\pi}=140$~MeV as input to obtain remaining parameters in the usual manner.   }
\label{table:NJLParamSet}
 \end{table*}
%

The present line of research complements our recent work 
by changing the model for hadron structure and this, in turn, may influence 
nuclear matter properties and hyperon optical potentials. Throughout we use the 
same notation and methods as in our earlier 
works~\cite{MyThesis,PhysRevC.89.065801,Whittenbury:2015ziz,ManuelsThesis,Carrillo-Serrano:2014zta}.

\section{QMC Model for Nuclear Matter}
\label{sec:QMC}
In our calculations we consider only the 
spin-$1/2$ octet baryons.
These interact via the exchange of mesons which
couple directly to the quarks. 
The exchanged mesons included are the scalar-isoscalar ($\sigma$),
vector-isoscalar ($\omega$), vector-isovector ($\rho$)  
and pseudo-scalar-isovector ($\pi$). 
These mesons only couple with the light quarks
by the phenomenological OZI rule. We include the 
full vertex structure  for the vector mesons,  
that is, both the Dirac and Pauli terms.

The QMC Lagrangian density used in this work is given by a combination
of baryon and meson components
\begin{equation}
\mathcal{L} = \sum_{B}\mathcal{L}_{B} 
+ \sum_m \mathcal{L}_{m}\ ,
\label{eq:wholeL}
\end{equation}
for the octet of baryons \mbox{$B \in
  \{N,\Lambda,\Sigma,\Xi\}$}  and selected
mesons \mbox{$m \in \{\sigma,\omega,\rho,\pi\}$}
with the individual Lagrangian
densities
\begin{eqnarray}
\mathcal{L}_{B} & = &\label{eq:lb}
\bar{\Psi}_{B} \bigg(
i\gamma_{\mu}\partial^{\mu} - M_{B} + g_{\sigma B}(\sigma)\sigma \nonumber \\
&&- g_{\omega B}\gamma^{\mu}\omega_{\mu} -\frac{f_{\omega B}}{2M_{\rm N}}\sigma^{\mu\nu}\partial_{\mu}\omega_{\nu} \nonumber \\
&&  - g_{\rho B}\gamma^{\mu}\boldsymbol{t}\cdot\boldsymbol{\rho}_{\mu} -\frac{f_{\rho B}}{2M_{\rm N}}\sigma^{\mu\nu}\boldsymbol{t}\cdot\partial_{\mu}\boldsymbol{\rho}_{\nu} \nonumber \\
&&-\frac{g_{\rm A}}{2f_{\pi}}\chi_{BB}\gamma^{\mu}\gamma^{5}\boldsymbol{\tau}\cdot\partial_{\mu}\boldsymbol{\pi}
 \bigg) \Psi_{B}\ ,
\end{eqnarray}
\begin{eqnarray}
\sum_{m}\mathcal{L}_{m} &=& \frac{1}{2}(\partial_{\mu}\sigma\partial^{\mu}\sigma 
- m_{\sigma}^{2}\sigma^{2}) \nonumber\\
&& - \frac{1}{4}\Omega_{\mu\nu}\Omega^{\mu\nu} 
+ \frac{1}{2}m_{\omega}^{2}\omega_{\mu}\omega^{\mu} \nonumber \\
&& 
- \frac{1}{4}\boldsymbol{R}_{\mu\nu}\cdot\boldsymbol{R}^{\mu\nu} 
+ \frac{1}{2}m_{\rho}^{2}\boldsymbol{\rho}_{\mu}\cdot\boldsymbol{\rho}^{\mu} \nonumber\\
&&+ \frac{1}{2}(\partial_{\mu}\boldsymbol{\pi}\cdot\partial^{\mu}\boldsymbol{\pi} 
- m_{\pi}^{2}\boldsymbol{\pi}\cdot \boldsymbol{\pi})\ ,
\label{eq:lm}
\end{eqnarray}
for which the vector meson field strength tensors are 
$\Omega_{\mu\nu}=\partial_{\mu}\omega_{\nu}-\partial_{\nu}\omega_{\mu}$
and
$\boldsymbol{R}_{\mu\nu}=\partial_{\mu}\boldsymbol{\rho}_{\nu}-\partial_{\nu}\boldsymbol{\rho}_{\mu}$.
{}$g_{iB}$, $f_{iB}$ denote the  meson-baryon coupling constants for 
the $i\in \left\lbrace \sigma , \omega , \rho\right\rbrace$ mesons.
The quantities in bold are vectors in isospin space, with isospin matrices 
denoted by $\boldsymbol{t}$ and isospin Pauli matrices 
by $\boldsymbol{\tau}$. For nucleons and 
cascade particles $\boldsymbol{t} =\frac{1}{2}\boldsymbol{\tau}$.  
The pion-baryon interaction used here is assumed to be described by an
SU(3) invariant Lagrangian with the mixing 
parameter $\alpha = 2/5$~\cite{RikovskaStone:2006ta}
from which the hyperon-pion coupling constants can be given in terms of
the pion nucleon coupling~\cite{RevModPhys.35.916,RikovskaStone:2006ta,Massot2008}.

From the Lagrangian given in Eq.~(\ref{eq:wholeL}), we use 
the Euler-Lagrange equations to obtain a system of coupled, 
non-linear partial differential equations for the quantum fields.
This is a difficult system of equations to solve and to make the problem tractable
a number of approximations are usually applied, including static, no sea and 
mean field approximations, which are implemented here. Following  
Refs.~\cite{Guichon:2006er,RikovskaStone:2006ta,Massot2008,HuToki:2010},
we decompose each meson field into two parts, a mean field 
part $\langle\phi \rangle$ and a fluctuation part $\delta\phi$, 
such that $\phi = \langle\phi\rangle +\delta\phi$. The equations of motion 
are then solved order by order. The fluctuation terms are small with respect 
to the mean field contribution, with the exceptions being 
the $\pi$ and $\rho$ meson fluctuations.
%
 \begin{table}[t] 
  \centering
  \resizebox{\columnwidth}{!}{
 \begin{tabular}{ccccccccc}
  \toprule[1.5pt]
 & $G_{\rm S}$ & $G_{\rm a}$  & $Z_{\left[\ell\ell\right]}$ & $Z_{[\ell s]}$ & $Z_{\left\lbrace \ell\ell\right\rbrace }$& $Z_{\left\lbrace\ell s\right\rbrace}$ & $Z_{\left\lbrace ss\right\rbrace}$&  \\
 & [GeV$^{-2}$] & [GeV$^{-2}$] & &  &  & &  & \\
\midrule
   & 7.65 &2.61  & 14.81  & 16.42 &3.56 & 3.93  & 4.28 & \\
   \bottomrule[1.5pt]
 \end{tabular}
 }
 \caption{Diquark couplings determined by fitting $M_{N}$ and $g_{A}$.}
\label{table:DiquarkParamSet}
 \end{table}
%
  \begin{table}[t] 
   \centering
   \resizebox{\columnwidth}{!}{
 \begin{tabular}{ccccccc}
  \toprule[1.5pt]
  & $M_{\left[\ell\ell\right]}$ & $M_{[\ell s]}$ & $M_{\left\lbrace \ell\ell\right\rbrace }$& $M_{\left\lbrace\ell s\right\rbrace}$ & $M_{\left\lbrace ss\right\rbrace}$ & \\
 & [MeV]& [MeV] & [MeV] & [MeV] & [MeV] & \\
\midrule
  & 679.18 & 848.71 & 1038.54 & 1170.67  & 1301.00 &  \\
   \bottomrule[1.5pt]
 \end{tabular}
 }
 \caption{Diquark masses determined by pole condition. }
\label{table:DiquarkMasses}
 \end{table}
%
\begin{table*}[t]
\resizebox{\textwidth}{!}{
\begin{tabular}{cccccccccccccccc}
\toprule[1.5pt]
&$c$  & $G_{\omega}$
& $M_{\rm N}$ & $M_{\Sigma}$  & $M_{\Xi}$ & $M_{\Lambda}$
& $g_{\sigma N}^{\rm NJL}$ &  $d$ 
& $\omega_{\Sigma}$ &$\omega_{\Xi}$ & $\omega_{\Lambda}$ 
& $\tilde{\omega}_{\Sigma}$ &$\tilde{\omega}_{\Xi}$ & $\tilde{\omega}_{\Lambda}$& \\
&[GeV]&[GeV$^{-2}$] & [GeV] & [GeV] & [GeV]& [GeV]&  &[GeV$^{-1}$] &  & & &  & & \\
 \midrule
& 1.141& 6.279 & 0.94 & 1.23222 & 1.32 & 1.118 & 12.9852 & 1.39786 & 0.528979 & 0.38203 & 0.769547 & 0.571791 & 0.415508 & 0.752602 \\
  \bottomrule[1.5pt]
\end{tabular}
}
\caption{Mass parametrisations for the spin\textendash$1/2$ baryon octet obtained by quadratic fits to NJL model nuclear matter calculation. Tabulated quantities are the $c$ parameter used in the static approximation, free baryon masses, the $\sigma$\textendash$N$ coupling, scalar polarisability, and weights.}
\label{table:NJLMassParams}
\end{table*}
%
\begin{table*}
    \centering
    \resizebox{0.9\textwidth}{!}{
\begin{tabular}{lcccccccc}
\toprule[1.5pt]
  \multirow{2}{*}{Model/} & \multirow{2}{*}{\ $g_{\sigma N}$\ } &\multirow{2}{*}{\ $g_{\omega N}$\ }&\multirow{2}{*}{\ $g_{\rho}$\ }& $K_{0}$&$L_{0}$ & 
  $U_{\Lambda}$ & $U_{\Sigma^{-}}$ & $U_{\Xi^{-}}$\\
 Scenario  & & & & [MeV]& [MeV] &[MeV]& [MeV] &[MeV]  \\[2mm]
 \midrule\\
 {\footnotesize Hartree }  & 9.65 & 6.8 & 8.54 &261 & 87  & -55 & -17 & -26 \\

%
{\footnotesize Standard }  & 8.29 & 8.36 & 4.92 & 263 & 81 & -5 & 27 & -5 \\  

{\footnotesize $\Lambda = 1.3$ }  & 8.55 & 9.48 & 5.24 & 278 & 84  & 16 & 49 & 5 \\

{\footnotesize Dirac Only }  & 9.41 & 7.95 & 7.66 & 277 &82 & -33 & 5 & -19\\

{\footnotesize $F_{\sigma}(\vec{k}) = 1$ }  & 8.86 & 8.11 & 4.24 & 259 & 75  & -22 & 14 & -14\\


\bottomrule[1.5pt]
    \end{tabular}
    }
 \caption{ Couplings, nuclear matter properties, 
and hyperon optical potentials determined for our standard 
case (for which $\Lambda=0.9$~GeV) and variations thereof. 
The symmetric nuclear matter quantities evaluated at saturation, 
$K_{0}$ and $L_{0}$, 
are the incompressibility and slope of the symmetry energy, 
respectively. The hyperon optical potentials 
are evaluated as in Ref.~\cite{MyThesis,PhysRevC.89.065801}.
\protect\label{table:NM}}  
\end{table*}

In the Fock terms a dipole form factor is used with a cut-off $\Lambda$. 
The same cut-off is used for all mesons. We consider several model variations,  
taking the cut-off $\Lambda = 0.9$~GeV as our ``Standard" or baseline scenario, 
which includes both Dirac (vector) and Pauli (tensor) interactions for 
the vector mesons. The other scenarios, which involve variations on 
the baseline are ``Hartree", which only includes the Hartree 
contribution; ``$\Lambda =1.3$~GeV", which has an increased 
cut-off; ``Dirac Only" which neglects the tensor contribution; 
and finally ``$F_{\sigma}(\vec{k})=1"$, where we take a hard form factor for 
the sigma meson, leaving the density dependence as determined within the model.

\section{Baryon Structure in the NJL model}
\label{sec:faddeev}
The NJL model~\cite{Nambu:1961tp,Nambu:1961fr} has been extensively studied, 
including a large number of reviews~\cite{Vogl:1991qt,Klevansky:1992qe,Hatsuda:1994pi,Ripka:1996fw,Ripka:1997zb,Buballa:2003qv}. 
Recently various phenomena related to hadron structure have been investigated 
using the NJL model with Schwinger's proper time regularisation 
modified in a manner that forbids the quarks to propagate on-mass-shell~\cite{Ebert:1996vx,Hellstern:1997nv}, 
in order to simulate quark
confinement~\cite{Cloet:2014rja,Carrillo-Serrano:2014zta,Carrillo-Serrano:2015uca,ManuelsThesis,Ninomiya:2014kja} and dimensional regularisation~\cite{Liu:2014vha}. 
In particular, the work of Ref.~\cite{Carrillo-Serrano:2014zta} is followed 
closely in the present calculation and subsequent 
parametrisation of the octet baryon masses.

We work with just the local (contact) four Fermi interaction between quarks,
which is parametrised by a coupling constant $G_\pi$ in the SU(3)-flavour 
NJL Lagrangian density. It is common to include a six-fermion term to describe 
phenomenologically the breaking of $U(1)$-axial symmetry, but as 
the $\eta$ and $\eta^\prime$ mesons play no role in the current work, we 
omit this term. The dynamic breaking of chiral symmetry is evident in 
the spontaneous generation of constituent quark 
masses ($M_u = M_d = M_\ell$ or $M_s$), which are determined by the 
so-called gap equation~\cite{Vogl:1991qt,Klevansky:1992qe,Hatsuda:1994pi,Ripka:1996fw,Ripka:1997zb,Buballa:2003qv}. 
The application of Fierz transformations to the NJL Lagrangian rearranges 
the fermion fields into meson and diquark channels. 
The resulting diquark Lagrangian density reads~\cite{Ishii:1995bu}
\begin{align}
\mathcal{L}_{I}^{qq} &= G_{\rm S} \Bigl[\bar{\psi}\,\gamma_5\, C\,\lambda_a\,\beta_A\, \bar{\psi}^T\Bigr]
\Bigl[\psi^T\,C^{-1}\gamma_5\,\lambda_a\,\beta_A\, \psi\Bigr] \nonumber \\
&
+ G_{\rm a} \Bigl[\bar{\psi}\,\gamma_\mu\,C\,\lambda_s\,\beta_A\, \bar{\psi}^T\Bigr]
\Bigl[\psi^T\,C^{-1}\gamma^{\mu}\,\lambda_s\, \beta_A\, \psi\Bigr] \, ,
\label{eq:qqlagrangian}
\end{align}
where $C$ corresponds to the charge conjugation matrix, 
which in our notation is $C=i\gamma_2\gamma_0$. Flavour is described by the 
usual SU(3) matrices $\lambda_a$, with $(a=2,5,7)$, and $\lambda_s$, 
with $(s=0,1,3,4,6,8)$, while the color $\bar{3}$ states are represented 
by $\beta_A = \sqrt{3/2}\lambda_A$, with $(A=2,5,7)$~\cite{Ishii:1995bu,Ishii:1993np,Ishii:1993rt}. 
This allows the description of effective $qq$ interactions in 
the scalar and axial-vector diquark channels, with strengths given by 
the coupling constants $G_{\rm S}$ and $G_{\rm a}$, respectively. 
%
\begin{figure}[tb]
\centering\includegraphics[width=\columnwidth,clip=true,angle=0]{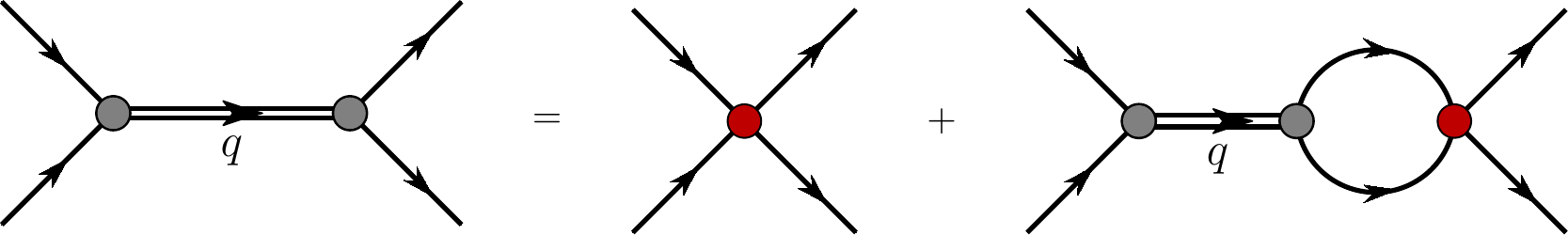}
\caption{(Color online) 
Inhomogeneous Bethe-Salpeter equation for diquark correlations.}
\label{fig:1}
\end{figure}

Diquarks are then described as $qq$ bound states through the solution of 
the Bethe-Salpeter equation depicted in Fig.~\ref{fig:1}. 
These solutions are given by the following reduced 
$t$-matrices\footnote{We follow the notation of 
Refs.~\cite{ManuelsThesis,Carrillo-Serrano:2014zta} where square brackets, 
$[q_1q_2]$, represent a scalar diquark with quark content $q_1$ and $q_2$, 
while $\{q_1q_2\}$ is the corresponding axial-vector diquark.}
\begin{align}
\label{eq:tscalar}
\tau_{[q_1q_2]}(q) 
&= \frac{4i\,G_{\rm S}}{1 + 2\,G_{\rm S}\,\Pi_{[q_1q_2]}(q^2)}, \\
\label{eq:taxial}
\tau^{\mu\nu}_{\{q_1q_2\}}(q) 
&= \frac{4\,i\,G_{\rm a}}{1 + 2\,G_{\rm a}\,\Pi^T_{\{q_1 q_2\}}(q^2)}\left(g^{\mu\nu} - \frac{q^\mu q^\nu}{q^2}\right) \nonumber \\
&
+ \frac{4\,i\,G_{\rm a}}{1 + 2\,G_{\rm a}\,\Pi^L_{\{q_1 q_2\}}(q^2)}\,\frac{q^\mu q^\nu}{q^2},
\end{align}
where the bubble diagrams read~\cite{ManuelsThesis,Carrillo-Serrano:2014zta}
\begin{align}
\label{eq:bubble_PP}
&\Pi_{[q_1q_2]}\left(q^2\right) = 6i \int \frac{d^4k}{(2\pi)^4}\ \mathrm{Tr}\left[\gamma_5\,S_{q_1}(k)\,\gamma_5\,S_{q_2}(k+q)\right], \\
\label{eq:bubble_VV}
&\Pi^T_{\{q_1q_2\}}(q^2)\left(g^{\mu\nu} - \frac{q^\mu q^\nu}{q^2}\right) + \Pi^L_{\{q_1q_2\}}\frac{q^\mu q^\nu}{q^2} = \nonumber \\
&\hspace{17mm}6i \int \frac{d^4k}{(2\pi)^4}\ \mathrm{Tr}\left[\gamma^\mu\,S_{q_1}(k)\,\gamma^\nu\,S_{q_2}(k+q)\right].
\end{align}
The traces are taken over Dirac indices only and $S_q(k)$ is 
the Feynman constituent quark propagator.
The pole positions of the $t$-matrices are the masses of the different 
scalar and axial-vector diquarks, $M_{[q_1q_2]}$ and $M_{\{q_1q_2\}}$, 
respectively. In addition, in the solution of the Faddeev equations we use the following pole approximations for the reduced $t$-matrices  
\begin{align}
\label{eq:scalarpropagatorpoleform}
\tau_{[q_1q_2]}(q) &\to  \frac{-i\,Z_{[q_1q_2]}}{q^2 - M_{[q_1q_2]}^2+ i\varepsilon} \, , \\
\label{eq:axialpropagatorpoleform}
\tau^{\mu\nu}_{\{q_1q_2\}}(q) &\to \frac{-i\,Z_{\{q_1q_2\}}}{q^2 - M_{\{q_1q_2\}}^2 + i\varepsilon} 
\left(g^{\mu\nu} - \frac{q^{\mu}q^{\nu}}{M_{\{q_1q_2\}}^2}\right) \, .
\end{align}
The residues at the poles define the effective couplings 
$Z_{[q_1q_2]}$ and $Z_{\{q_1q_2\}}$~\cite{ManuelsThesis,Cloet:2014rja,Carrillo-Serrano:2014zta}.
%
\begin{figure}[tb]
\centering\includegraphics[width=\columnwidth,clip=true,angle=0]{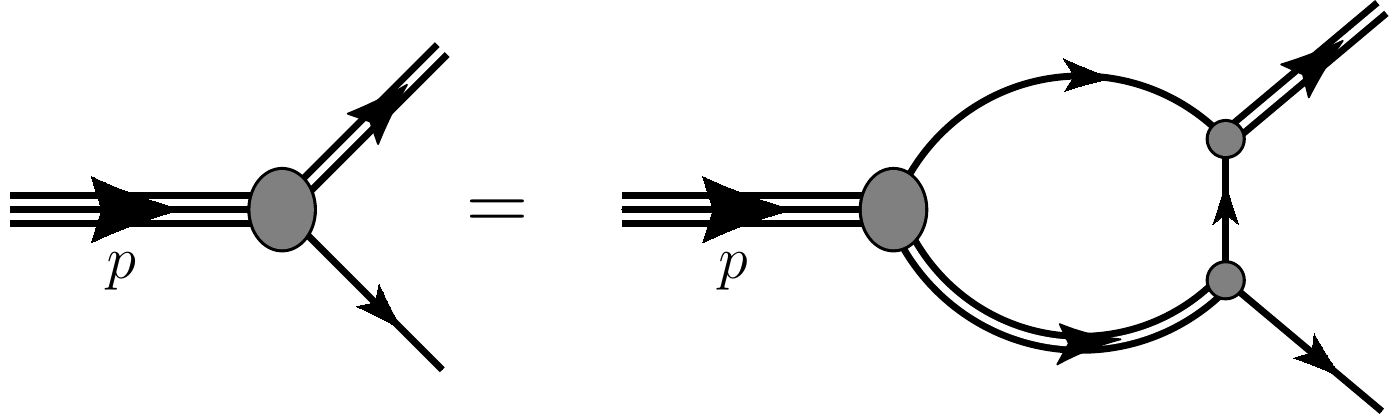}
\caption{Homogeneous Faddeev equation for the each member of 
the baryon octet. The masses and 
vertices are found from its solution.}
\label{fig:faddeev_equation}
\end{figure}

{}For each member of the baryon octet we solve the Faddeev 
equations~\cite{Afnan:1977pi} in the 
quark-diquark picture, depicted graphically in Fig.~\ref{fig:faddeev_equation}.
The formalism and the solution of the Faddeev equations in the present 
NJL model is detailed in Refs.~\cite{ManuelsThesis,Carrillo-Serrano:2014zta}. 
The analytic form of the homogeneous Faddeev equations is given by
\begin{equation}
\Gamma_B(p,s) = Z_B\Pi_B(p)\Gamma_B(p,s),
\label{eq:faddev_equation}
\end{equation}
which amounts to a homogeneous Fredholm equation of the second 
kind for each baryon, $B$.
\begin{figure}[tb]
\centering\includegraphics[width=\columnwidth,clip=true,angle=0]{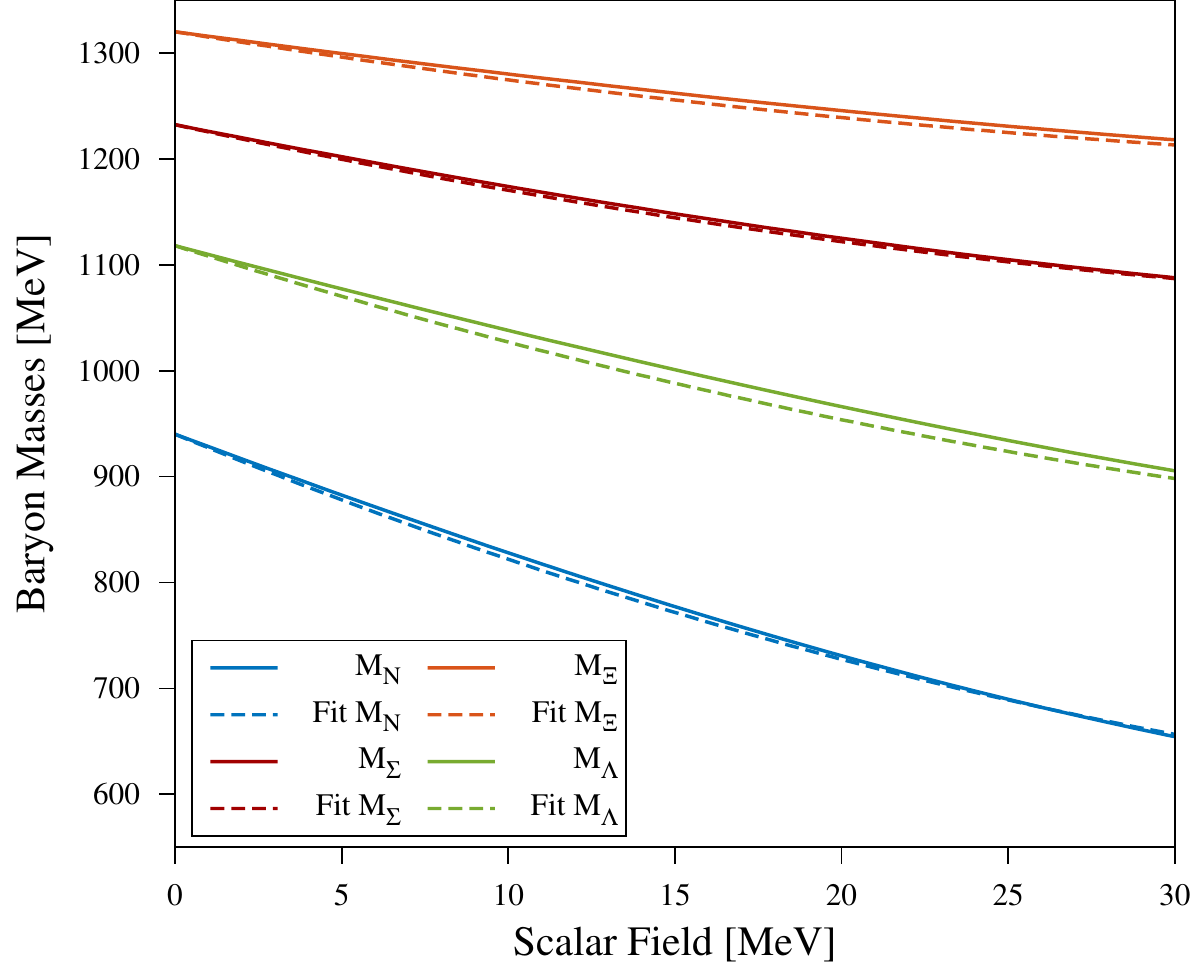}
\caption{(Color online) Baryon octet masses vs scalar field. 
The data from the solution of the Faddeev equation is shown as a continuous 
line and the fit from Eq.~\eqref{eq:massparam}.}
\label{fig:mass_param}
\end{figure}
{}Following Ishii {\em et al.} we employ the static 
approximation~\cite{Ishii:1993np,Buck:1992wz} for the Feyman propagator 
of the exchanged quark. The formulation of the Faddeev kernel for each member 
of the octet ($B$) uses the quark-diquark bubble diagrams, $\Pi_B(p)$, 
and the normalised Faddeev vertices, $\Gamma_B(p,s)$. 
The explicit form of the Faddeev vertices, bubble matrices and 
quark exchange kernels are explained in detail 
in Refs.~\cite{ManuelsThesis,Carrillo-Serrano:2014zta}. 
By solving the Faddeev equations we  obtained the masses of the octet 
baryons. Note that the nucleon and cascade masses in vaccuum 
($M_N$ and $M_\Xi$) were used as input to constrain $G_{\rm a}$ 
and $M_s$~\cite{ManuelsThesis,Carrillo-Serrano:2014zta}.

The free parameters of our model are the infra-red cut-off and the light constituent quark mass, which we take to be $\Lambda_{IR} = 240$ MeV and $M_\ell = 400$ MeV respectively. The infra-red cut-off is of the order of $\Lambda_{QCD}$, because it crudely simulates quark confinement~\cite{Ebert:1996vx,Hellstern:1997nv,Bentz:2001vc}. For the other parameters we fit them in the following way: $\Lambda_{UV}$ and $G_\pi$ are fit to reproduce the experimental values of the pion's decay constant and its mass, while $G_{\rm S}$ is chosen to obtain the experimental axial charge $g_A$ of the nucleon. The computation of $g_A$ is done following the formalism of Ref.~\cite{ManuelsThesis,Carrillo-Serrano:2014zta}.

The values of the baryon octet masses obtained are summarised 
in Table~\ref{table:NJLMassParams}. Following the work of 
Bentz and Thomas~\cite{Bentz:2001vc}, to account for the lack of agreement 
between the static approximation and the exact result of the Faddeev equations 
in nuclear matter, the exchange quark propagator is replaced by the following 
interpolating function
\begin{equation}
\frac{1}{M_\ell^*} \rightarrow \frac{1}{M_\ell}\frac{M_\ell + c}{M_\ell^* + c},
\label{eq:q_prop_interpolation}
\end{equation}
where $M_\ell^*$ is the value of the constituent quark mass for a baryon 
in nuclear matter and $M_\ell$ is its value inside a free baryon. 
$c$ is set to 1.141 GeV and with the $\omega N$
coupling, $G_{\omega}$, set as in Table~\ref{table:NJLMassParams}, we reproduce 
the saturation properties of symmetric nuclear matter in the NJL model.
We check that a variation of $c$ from $0.5$~GeV to $2.5$~GeV makes no significant 
change in the results quoted here.

In the NJL model, the scalar field is related to the scalar potential 
$\Phi = M_\ell^* - M_\ell$ by
\begin{equation}
\bar{\sigma} = \frac{\Phi}{\sqrt{Z_{\pi}(0)}}\, ,
\label{eq:NJL_scalar_field}
\end{equation}
where $Z_{\pi}(0)$ is defined, in analogy to Eq.~(\ref{eq:scalarpropagatorpoleform}) and (\ref{eq:axialpropagatorpoleform}), as the residue at the pole in the pion t-matrix.
\begin{figure}
\centering
\includegraphics[width=\linewidth]{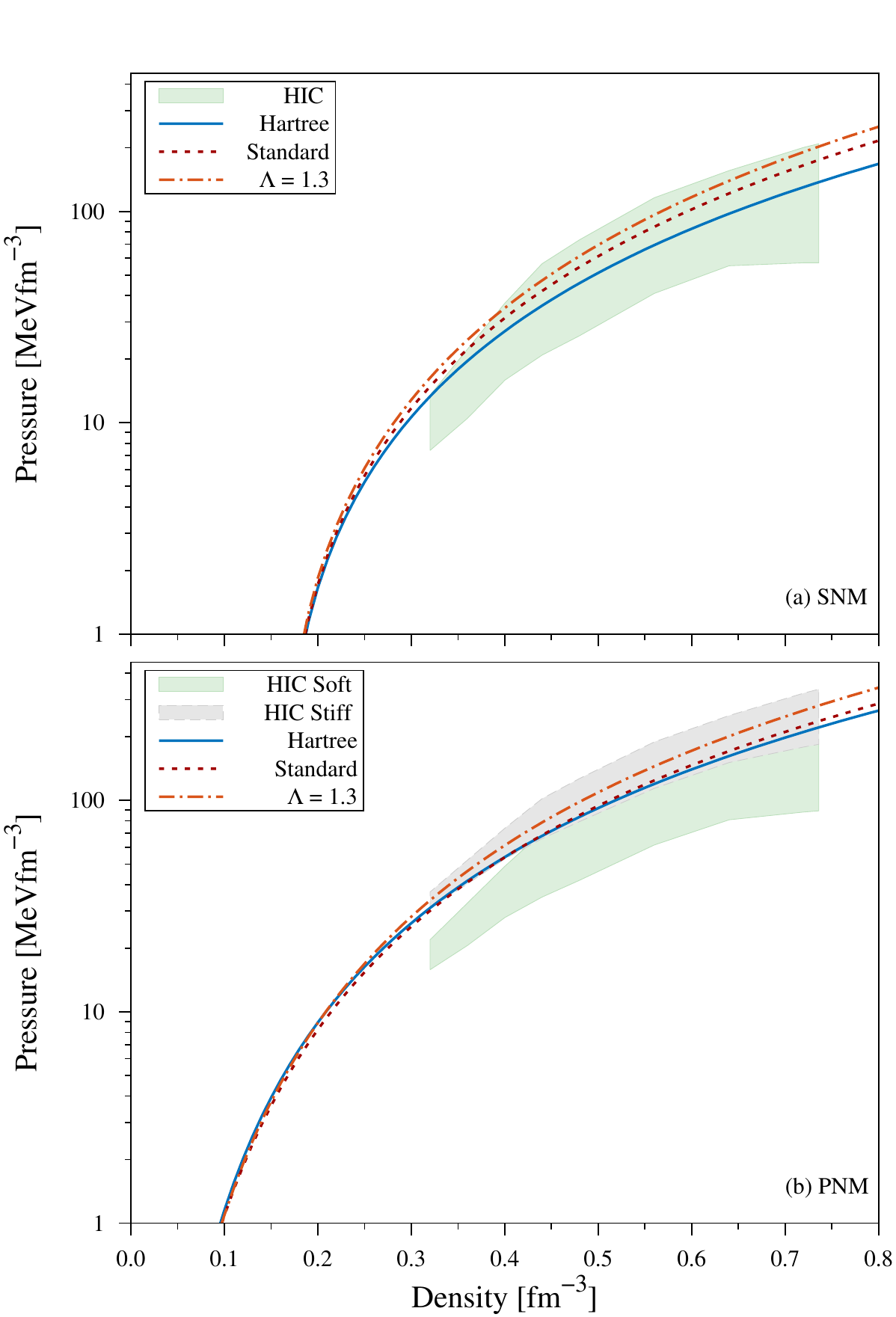} 
\caption{Pressure as a function of density in (a) SNM and (b) PNM. 
The constraints come from heavy-ion collision experiments 
deduced in Ref.~\cite{Danielewicz:2002pu}.}
\label{fig:PVsDen}
\end{figure}
Changes in the scalar field are linked to variations of the constituent 
quark masses (inside the baryons and diquarks) in nuclear matter. 
Consequently, the Faddeev equations were solved as a function of the 
in-medium constituent quark masses. The baryon masses were then  
parametrised as functions of mean scalar field, $\bar{\sigma}$, 
using Eq.~\eqref{eq:massparam}~and~\eqref{eq:NJL_scalar_field}.
These parametrisations are given in Table~\ref{table:NJLMassParams}.
In Fig.~\ref{fig:mass_param} we show the calculated baryon masses as 
a function of the mean scalar field, together with the fits obtained 
assuming the form given by Eq.~\eqref{eq:massparam}.  
The range of field strength has been chosen to correspond to the range 
of densities explored in Fig.~\ref{fig:PVsDen}.
It is self-evident that the fits are in very good agreement with 
the calculated solution of the Faddeev equations.

\section{Nuclear Matter}
We now present the numerical results for the properties of nuclear matter 
obtained using the self-consistent solution of the NJL model described earlier.
To be definite, what we actually use from the mass parametrisations 
is the value of the scalar polarisability, $d$, and the weights, 
$\omega_{B}$ and $\tilde{\omega}_{B}$. The vacuum contribution that 
would normally be included in a quark matter calculation of 
the EoS~\cite{MyThesis,Whittenbury:2015ziz} is omitted for hadronic matter.

Table~\ref{table:NM} contains coupling constants, nuclear matter properties 
(incompressibility $K_{0}$ and slope of the symmetry energy $L_{0}$ at saturation)  
and hyperon optical potentials. 
The pressure as a function of density for SNM and PNM are shown 
in Fig.~\ref{fig:PVsDen} in comparison with constraints from heavy-ion collisions.
We find a slightly softer EoS when using the NJL mass parametrisations, 
than what we found using the bag model parametrisations 
in Refs.~\cite{MyThesis,PhysRevC.89.065801}. This is clearly illustrated, 
for example, by the value of the incompressibility at saturation and in 
the behaviour of the pressure as a function of density,
shown in  Fig.~\ref{fig:PVsDen}.

The incompressibility and hyperon optical potentials show only a minor
dependence on the $c$ parameter, which was introduced to handle 
the quark exchange in a simplified manner. The incompressibility reduces by 
just $8$\textendash $18$~MeV when $c$ is increased from $0.5$ to $2.5$~GeV. 
The $\Sigma^{-}$ optical potential is the most sensitive of the optical potentials 
to a  variation of $c$, exhibiting a reduction of $11$\textendash $16$~MeV, 
over the same range. 

As the hyperon optical potentials are determined within the model without 
readjustment to empirically determined values, we find it encouraging  
to discover a reasonable level of agreement between several model variations 
and the empirically determined values. Of particular interest is the scenario 
which deviates from our standard scenario by the use of a hard 
scalar form factor. The motivation for taking the hard form factor only for 
the sigma meson is that its coupling already includes a density 
dependence obtained through our model of hadron structure,
which naturally acts to reduce the scalar Fock term at high density.

\section{Summary}
We have self-consistently solved for the structure of the octet baryons 
imbedded in nuclear matter, using the NJL model as the underlying model of 
hadron structure. Using those solutions we have presented numerical 
results for Symmetric Nuclear Matter (SNM), Pure Neutron Matter (PNM) and 
the hyperon optical potentials. Overall the results are very reasonable, 
with the properties of both SNM and PNM in good agreement with heavy ion 
constraints over the entire range up to five times nuclear matter density.
{}For most of the scenarios explored the $\Lambda$ and $\Xi$ are bound by 
reasonable amounts, while the $\Sigma$ is unbound, as suggested by phenomenological 
studies.

With a view to future applications to neutron star structure, we observe that 
the EoS is softer than what we obtained with the bag model parametrisations 
in our earlier work.  This may well lead to a somewhat lower maximum mass, 
unless there is a transition to quark matter.
In the near future we will explore the consequences of this model for neutron star 
properties, with and without such a transition. In terms of theoretical 
improvements it would clearly be valuable to move beyond the static approximation, 
making an exact solution with the full exchanged-quark propagator.

\section*{Acknowledgements}
This work was supported by the University of Adelaide and by the Australian 
Research Council through the ARC Centre of Excellence for Particle Physics
at the Terascale CE110001004, an ARC Australian Laureate Fellowship FL0992247 
and an ARC Discovery Project DP150103101.

  \bibliographystyle{elsarticle-num} 





\bibliography{QMC_NJL_NM_paper}
\bibliographystyle{unsrt}
\end{document}